\documentclass[showpacs,prb]{revtex4}
\usepackage{graphicx}
\begin{document}
\title{Substrate-induced strain effects on Pr$_{0.6}$Ca$_{0.4}$MnO$_3$ films\\}

\author{C.S. Nelson}
\affiliation{NRL-SRC, National Synchrotron Light Source, Brookhaven National Laboratory, Upton, NY  11973-5000}
\author{J.P. Hill}
\affiliation{Department of Physics, Brookhaven National Laboratory, Upton, NY  11973-5000}
\author{Doon Gibbs}
\affiliation{Department of Physics, Brookhaven National Laboratory, Upton, NY  11973-5000}
\author{M. Rajeswari}
\affiliation{Department of Physics, Astronomy, and Geosciences, Towson University, Towson, MD  21252}
\author{A. Biswas}
\affiliation{Center for Superconductivity Research, Department of Physics, University of Maryland, College Park, MD  20742}
\author{S. Shinde}
\affiliation{Center for Superconductivity Research, Department of Physics, University of Maryland, College Park, MD  20742}
\author{R.L. Greene}
\affiliation{Center for Superconductivity Research, Department of Physics, University of Maryland, College Park, MD  20742}
\author{T. Venkatesan}
\affiliation{Center for Superconductivity Research, Department of Physics, University of Maryland, College Park, MD  20742}
\author{A.J. Millis}
\affiliation{Department of Physics, Columbia University, New York, NY  10027}
\author{F. Yokaichiya}
\affiliation{Instituto de Fisica, Universidade Estadual de Campinas, CP 6165, Campinas, SP, Brazil}
\author{C. Giles}
\affiliation{Instituto de Fisica, Universidade Estadual de Campinas, CP 6165, Campinas, SP, Brazil}
\author{D. Casa}
\affiliation{CMC-CAT, Advanced Photon Source, Argonne National Laboratory, Argonne, IL  60439}
\author{C.T. Venkataraman}
\affiliation{CMC-CAT, Advanced Photon Source, Argonne National Laboratory, Argonne, IL  60439}
\author{T. Gog}
\affiliation{CMC-CAT, Advanced Photon Source, Argonne National Laboratory, Argonne, IL  60439}

\date{\today}
\begin{abstract}
We report the characterization of the crystal structure, low-temperature charge and orbital ordering, transport, and magnetization of Pr$_{0.6}$Ca$_{0.4}$MnO$_3$ films grown on LaAlO$_3$, NdGaO$_3$, and SrTiO$_3$ substrates, which provide compressive (LaAlO$_3$) and tensile (NdGaO$_3$ and SrTiO$_3$) strain.  The films are observed to exhibit different crystallographic symmetries than the bulk material, and the low-temperature ordering is found to be more robust under compressive--- as opposed to tensile--- strain.  In fact, bulk-like charge and orbital ordering is not observed in the film grown on NdGaO$_3$, which is the substrate that provides the least amount of nominal and measured, but tensile, strain.  This result suggests the importance of the role played by the Mn--O--Mn bond angles in the formation of charge and orbital ordering at low temperatures.  Finally, in the film grown on LaAlO$_3$, a connection between the lattice distortion associated with orbital ordering and the onset of antiferromagnetism is reported.
\end{abstract}
\pacs{75.47.Gk, 68.55.Jk, 78.70.Ck, 73.50.-h, 75.70.Ak}
\maketitle

\section{Introduction}

Manganite films have been of interest since the first observation of the colossal magnetoresistance (CMR) effect in a perovskite-type La$_{0.67}$Ca$_{0.33}$MnO$_3$ film.\cite{jin}  The enhanced effect of the change in resistivity upon application of a magnetic field--- compared to that observed in the more well-known giant magnetoresistance materials--- immediately suggested device applications based on the sensitivity to magnetic fields.  More recently, ferroelectric field effect,\cite{mathews} bolometric,\cite{goyal} optical,\cite{kida} and spintronic \cite{wolf} devices, which take advantage of additional properties exhibited by some manganite films, have also been proposed.  

For all manganite film applications, a key parameter in the behavior of the system is the substrate-induced strain caused by the lattice mismatch between the manganite material and the substrate.  Substrate-induced strain has been reported to change a wide variety of properties, with examples including the crystal symmetry,\cite{koo,zandbergen,venimadhav,song} transport \cite{prellier,klein} and magnetic \cite{wu} anisotropies, the magnitudes of ferromagnetic T$_c$ \cite{millis} and the melting field,\cite{prellier2,yang} the spin and orbital order structure,\cite{izumi} and the tendency toward phase separation.\cite{biswas}  Theoretical work also predicts substrate-induced strain effects on the properties exhibited by manganite films.  For example, a 2\% tensile strain is expected to alter the antiferromagnetic ground state of LaMnO$_3$ from ($\pi\pi$0)- to ($\pi\pi\pi$)-type ordering.\cite{ahn}  

The influence that substrate-induced strain appears to exhibit over so many properties of manganite films suggests that it could be used advantageously--- in order to enhance the property of interest in a given material.  That is, substrate-induced strain could be used to tune the behavior of the films.  But before such tuning becomes more than a possibility, the role of substrate-induced strain in these materials needs to be more clearly understood.  In a step toward this goal, we report a study of the effects of substrate-induced strain on Pr$_{0.6}$Ca$_{0.4}$MnO$_3$--- a CMR material for which the bulk properties have been thoroughly investigated,\cite{jirak,tomioka,zim_prl,zim_to,john,zim_long,daoud} and which is therefore ideal for investigations of the effects of substrate-induced strain.  The type of questions that we hope to answer are:  how does substrate-induced strain affect the type of low-temperature ordering, the ordering temperature, and the correlation length?  And how do these results connect to transport and magnetization in the films?  In order to address these questions, we have studied three Pr$_{0.6}$Ca$_{0.4}$MnO$_3$ films grown on LaAlO$_3$ (LAO), NdGaO$_3$ (NGO), and SrTiO$_3$ (STO) substrates.  The results of this study are that both compressive (film grown on LAO) and tensile (films grown on NGO and STO) strain result in films with a different room temperature, crystallographic symmetry than is observed in the bulk material.  We also observe bulk-like charge and orbital ordering--- with a reduced ordering temperature and a shorter correlation length--- in the film grown on LAO.  In addition, we find that tensile strain has a stronger effect on the low-temperature ordering, and even a 0.3\% measured, tensile strain appears to inhibit the formation of bulk-like charge and orbital order in the film grown on NGO.  Finally, in the film grown on LAO, we observe a connection between the lattice distortion associated with orbital ordering and the onset of antiferromagnetic ordering.

The paper is organized into several sections.  We begin by briefly describing the bulk material in section II, and then move on to reporting the results of studies carried out on three films with compressive, and different amounts of tensile, substrate-induced strain.  These latter results are grouped into sections on the crystal structure (III), low-temperature ordering (IV), and transport and magnetization (V).  Finally, we conclude with a summary of our results in section VI.

\section{Bulk P\lowercase{r}$_{0.6}$C\lowercase{a}$_{0.4}$M\lowercase{n}O$_3$}

Pr$_{0.6}$Ca$_{0.4}$MnO$_3$ has a perovskite structure, with {\it Pbnm} symmetry at room temperature.\cite{jirak}  It exhibits the so-called GdFeO$_3$ distortion due to the tilting of the MnO$_6$ octahedra, as well as a cooperative Jahn-Teller distortion at temperatures below $\sim$245 K.  Associated with this latter distortion is believed to be a complicated charge and orbital ordering pattern, which consists of a checkerboard arrangement of inequivalent Mn sites that are conventionally referred to as Mn$^{3+}$ and Mn$^{4+}$.\cite{dispro}  These planes are stacked in phase along the {\it c}-axis.  In addition, there is a doubling of the unit cell along the {\it b}-axis that is commonly associated with an alternating pattern of occupied {\it 3d$_{z^2-r^2}$}-type orbitals of Mn$^{3+}$ ions.  At lower temperatures, below $\sim$140 K, Pr$_{0.6}$Ca$_{0.4}$MnO$_3$ orders magnetically in a structure consisting of ferromagnetic zigzags, coupled antiferromagnetically.  And finally, below $\sim$40 K, the magnetic moments are believed to become canted with a net moment along the {\it c}-axis,\cite{tomioka} which can be thought of as a compromise between the ordering exhibited at a doping of x = 0.5 (moments in the {\it a-b} plane, and stacking antiparallel along the {\it c}-axis) and x = 0.3 (moments lying--- and stacking parallel--- along the {\it c}-axis).  

Recently, resonant x-ray scattering--- in which the incident photon energy is tuned near the Mn K absorption edge--- has been shown to be sensitive to the presence of charge and orbital order,\cite{murakami} such as exhibited in Pr$_{0.6}$Ca$_{0.4}$MnO$_3$.  With this technique, the scattering intensity at charge and orbital ordering wavevectors is signficantly enhanced, which enables high-resolution, quantitative studies of these cooperative ordering phenomena.  Resonant scattering from ordered systems also exhibits polarization- and azimuthal-dependent properties that provide additional information about the details of the ordering.  In bulk Pr$_{0.6}$Ca$_{0.4}$MnO$_3$, resonant x-ray scattering has been used to verify the type of orbital order--- described above--- and measure the temperature-dependence of both the charge and orbital order correlation lengths.\cite{zim_prl,zim_long}  From the latter measurements, the charge order was observed to drive the orbital order, and the orbital order was found to be short-ranged at all temperatures below the transition temperature (T$_{co}$), reaching a maximum correlation length of 320 $\pm$ 10 $\AA$ at low temperatures.  In addition to the resonant scattering observed at both the charge and orbital order wavevectors, nonresonant scattering was also reported and was attributed to the presence of lattice distortions--- with both transverse and longitudinal components--- with the same periodicities as the charge and orbital order.\cite{zim_long}  These low-temperature lattice distortions indicate that the material undergoes a change in crystallographic symmetry, and other space groups have recently been proposed.\cite{daoud} 

\section{Room Temperature, Crystal Structures of Films}

A commercially-obtained, polycrystalline pellet was used to grow the 2500--3000 $\AA$ thick, Pr$_{0.6}$Ca$_{0.4}$MnO$_3$ films with the pulsed laser deposition method at the University of Maryland.\cite{rutherford}  LAO, NGO, and STO substrates were chosen in order to provide compressive (LAO) and different amounts of tensile (NGO and STO) strain to the films (see Table I).  The crystal structures of the films were investigated using x-ray diffraction on beamline X22C at the National Synchrotron Light Source and the CMC-CAT beamline 9-ID at the Advanced Photon Source.  Both beamlines are equipped with double-crystal monochromators (Ge(111) and Si(111), respectively) and 6-circle diffractometers in vertical scattering geometries.  For all measurements reported here, a graphite analyzer was used. 

The substrate surface-normal orientations of LAO(100), NGO(110), and STO(100) were observed to result in Pr$_{0.6}$Ca$_{0.4}$MnO$_3$ films with (110) or (002) orientations, in {\it Pbnm} notation.  Because of the similar values of {\it a}, {\it b}, and {\it c}/$\sqrt{2}$--- as well as the large diffraction widths of the Bragg peaks, which would obscure the presence of any crystallographic twins--- the particular orientations of the films remain ambiguous.  For simplicity in notation, we assume that the films are (110)-oriented in what follows.  The mosaic widths of the films, as characterized at the (220) Bragg peaks, were 0.2$^{\circ}$, 0.1$^{\circ}$, and 0.1--0.3$^{\circ}$ full-width-at-half-maximum for the films grown on LAO, NGO, and STO,\cite{3headed} respectively. 

For the films grown on LAO and NGO, clear evidence of a crystal symmetry lower than orthorhombic was observed at room temperature.  This evidence is displayed in Figure 1, which shows longitudinal scans through the (220) and (200) Bragg peaks.  The scans are plotted in reciprocal lattice units of the bulk material, and along ({\it H}00) and (0{\it K}0) (not shown) are found to peak at near-integer values.  In contrast, for scans along ({\it HH}0), which corresponds to the direction of the surface normal, the peaks are shifted away from the positions expected assuming the bulk, orthorhombic symmetry.  This shifting indicates that $\gamma$--- the angle between the {\it a} and {\it b} axes--- deviates from 90$^{\circ}$.  Specifically, in the film on LAO (NGO), the component of substrate-induced strain on {\it a} and {\it b} is compressive (tensile), which results in the two axes being pushed together (pulled apart) and therefore $\gamma$ is less (greater) than 90$^{\circ}$.

In order to determine the lattice parameters at room temperature, the positions of six peaks were measured.  This enabled a determination of the values of {\it a}, {\it b}, {\it c}, and $\gamma$, while initially assuming a monoclinic symmetry for all films (note that in the film grown on STO, $\gamma$ was measured to be 90$^{\circ}$, and the film has a higher symmetry--- tetragonal--- than either the two other films or the bulk material).  The measured values are displayed in Table II.  Note that these values are the average lattice constants of the films, and it is likely that the atomic spacings and microstructure vary throughout the films.  For example, it is possible that the film layers closest to the substrate are contracted (expanded) under compressive (tensile) strain--- in order to maintain perfect epitaxy with the substrate--- and that this strain is relaxed for film layers further away from the substrate.  As we shall discuss below, such relaxation is most dramatically in evidence in the film grown on STO.

In all three films, the measured values of the {\it a} and {\it b} lattice constants are observed to be close to the bulk values, while the in-plane lattice constants, {\it c}, follow the substrate.  As a measure of this latter effect, note the experimentally observed values of the substrate-induced strain displayed in Table I.  In this table, we have defined the nominal strain as the difference between the in-plane lattice constants of the substrates and the {\it c} lattice constant of bulk Pr$_{0.6}$Ca$_{0.4}$MnO$_3$.  Note that the nominal strain would be the strain experienced by the films if they were perfectly lattice-matched to the substrates.  The other gauge of substrate-induced strain listed in Table I--- the measured strain--- is defined as the difference between the observed, {\it c} lattice constants of the films and the {\it c} lattice constant of the bulk material, and is the actual strain experienced by the films.  The close correspondence between the nominal and measured strain is especially pronounced for the films grown on LAO and NGO.  In contrast, for the film grown on STO, a significant lattice mismatch between the film and the substrate exists.  In this film, there is also evidence of relaxation of the strain in longitudinal scans through ({\it HH}0) peaks, in that these peaks are extremely broad and have three maxima.  Since only two peaks can arise from twinning, the shape of the ({\it HH}0) peaks suggests that there is a range of out-of-plane lattice constants that is due to a relaxation of the strain within the film.  We note that such relaxation has been observed via x-ray microdiffraction measurements in La$_{1-x}$Sr$_x$MnO$_3$ films that were also grown under tensile strain.\cite{soh}

To test for different thermal expansion coefficients, which would result in temperature-dependent strain, the Bragg peak intensities and widths were also measured as a function of temperature.  The results from fits to these data for the films on LAO and STO are displayed in Figure 2.  While both the intensity and width of the (200) Bragg peak of the film grown on STO remain nearly constant as a function of temperature, the peak changes dramatically for the film grown on LAO.  Between 100 and 300 K, the correlation length, which is inversely proportional to the half-width-at-half-maximum (HWHM), nearly doubles.  These results indicate that the film grown on LAO is affected by the difference in thermal expansion coefficients between it and the substrate, and that it is more strained at low temperatures.

\section{Low-Temperature Ordering in Films}

The presence of charge and orbital ordering--- such as observed in the bulk material--- was investigated in each film at low temperatures.  This search was aided by the fact that in the bulk material, scattering associated with charge and orbital ordering occurs at structurally forbidden reflections.  Using the bulk, orthorhombic notation for which the Bragg peaks occur at (0 2{\it K} 0), peaks associated with charge and orbital order occur at (0 2{\it K}+1 0) and (0 {\it K}+$\frac{1}{2}$ 0), respectively, for integer {\it K}. 

\subsection{compressive:  on LAO}

The Pr$_{0.6}$Ca$_{0.4}$MnO$_3$ film grown on LAO was cooled to the base temperature of a closed-cycle cryostat ($\sim$10 K) and then heated to 100 K before being exposed to the incident x-ray beam--- in order to avoid possible x-ray induced melting of the ordering that is known to occur in bulk Pr$_{1-x}$Ca$_{x}$MnO$_3$, for {\it x} near 0.3.\cite{valery}  At a temperature of 100 K, peaks at the charge and orbital order wavevectors for the bulk material were observed, and were found to exhibit a resonant enhancement upon tuning the incident photon energy near the Mn K edge.  The (300) charge order peak was observed to have a width of similar magnitude to that of the nearby Bragg peaks, while the (2.5 0 0) orbital order peak was significantly broadened.\cite{twins}  Using a Lorentzian-squared lineshape to fit the data and defining the correlation length as $\textstyle\xi\equiv\frac{a}{2\pi\Delta H}$, where {\it a} is the lattice constant and $\Delta H$ is the HWHM value of the fit to the scan along {\it H}, the orbital order correlation length was determined to be 90 $\pm$ 10 $\AA$. 

Longitudinal scans through the (2.5 0 0) and (300) peaks were carried out as a function of temperature, and the fitted peak intensities are shown in Figure 3.  The (2.5 0 0) peak was observed to disappear at a temperature below 200 K.  In contrast, the (300) peak was observed to reach a minimum in intensity at this temperature and then {\it increase} in intensity with increasing temperature.  This intriguing behavior at the charge order wavevector was accompanied by a dramatic change in the energy dependence of the scattering, which can be seen in the insets displayed in Figure 3.  While the (300) peak intensity exhibits a maximum at an incident photon energy of $\sim$6.555 keV at a temperature of 100 K, the resonant enhancement decreases with increasing temperature and is absent at a temperature of 240 K.

Putting the results together, the film on LAO appears to exhibit the same charge and orbital ordering as in the bulk material, but with a reduced ordering temperature and shorter orbital order correlation length.  Another difference between the film and the bulk material is that while the nonresonant scattering associated with the charge ordering is observed to disappear at the transition temperature in the bulk material, it remains present at this wavevector even after the disappearance of the ordering in the film.  As a result, we do not associate the nonresonant scattering with the charge order, but rather as a further indication that the film and bulk material have different crystallographic symmetries, as was already noted in section III.  That is, the (300) peak can be thought of as a structural Bragg peak--- a conclusion that is further supported by a comparison of the high-temperature behaviors of the (300) and (200) peaks (see Figure 2), in which both peaks are observed to increase in intensity with increasing temperature.  For the (200) Bragg peak, we argued that this behavior was due to a difference in thermal expansion coefficients between the film and the substrate that resulted in the film being more strained at low temperatures, and the same explanation would apply to the behavior of a structural (300) peak.  We also note that temperature-dependent strain could also be at least partially responsible for the reduced orbital order correlation length of the film compared to that observed in the bulk material.

Scattering at the orbital order wavevector was also studied at a second point in reciprocal space--- at (2 1.5 0).  The peak intensity at this Q was approximately 50 times stronger than that at (2.5 0 0), which is presumably due to the lattice distortion being primarily transverse to the orbital order wavevector (i.e., the scattering intensity varies as $(\vec{Q}\cdot\vec{\delta})^{2}$, so the transverse component of the lattice distortion does not contribute to the scattering at (2.5 0 0)).  At (2 1.5 0), possible x-ray illumination effects were investigated at a temperature of 10 K.  No such effects were observed, and temperature-dependent data were collected over an extended range relative to the results described above.  These measurements, which show scattering consistent with the temperature-dependent behavior of the scattering at (2.5 0 0), will be discussed in subsection IV.C.

\subsection{tensile:  on NGO and STO}

In both of the Pr$_{0.6}$Ca$_{0.4}$MnO$_3$ films grown on substrates that provide tensile strain, peaks at the charge order wavevector for the bulk material were observed.\cite{caveat}  However, these peaks did not exhibit a resonant enhancement at low temperatures, had widths similar in magnitude to nearby Bragg peaks, and were found to persist up to the highest temperatures investigated ($>$600 K in the film grown on NGO, and 800 K in the film grown on STO).  The similarity in behavior between these peaks and Bragg peaks suggests that they are also allowed structural peaks--- and the result of the different crystallographic symmetry of the films compared to the bulk material--- as was also concluded for the film grown on LAO.

In the film grown on NGO, broad peaks with no evident resonant enhancement were observed at the orbital order wavevector for the bulk material at low temperatures.  Reciprocal space scans as a function of temperature through one of these peaks--- (2 1.5 0)--- are shown in Figure 4.  The peak is observed to disappear in the temperature range of 200--220 K, which is less than the transition temperature of the bulk material, and the low-temperature correlation length was measured to be 40 $\pm$ 10 $\AA$.  In addition to this commensurate peak, a hint of peaks with an incommensurability, $\epsilon$, of $\sim$0.04 rlu was observed.  These peaks can be most clearly seen at a temperature of 40 K, as shown in Figure 4.  Unfortunately, the extremely weak intensities of these possible peaks make it impossible to extract temperature-dependent information about them.

An unsuccessful search for peaks with the commensurate orbital ordering wavevector along (0{\it K}0) was also carried out for the film grown on NGO.  Based on the weak scattering intensity observed for peaks near the (220) Bragg peak, this failure is perhaps not surprising.  That is, any scattering at (0 2.5 0) would be expected to be prohibitively weak to observe--- assuming that the lattice distortion is primarily transverse to the orbital order wavevector, as is believed to be the case in both the bulk material and the film grown on LAO (see subsection IV.A., above).

In the film grown on STO, no peaks at the orbital ordering wavevector for the bulk material were observed near either the (220) or (020) Bragg peaks.  Wide scans over regions in reciprocal space that would include peaks due to incommensurate orbital ordering were also carried out, but no peaks were observed.

\subsection{low-temperature, charge and orbital ordering summary}

The film grown on STO, which is subject to the most, nominal substrate-induced strain, exhibits no low-temperature, charge or orbital ordering.  The nonuniform strain distribution suggested by scans in which Q is parallel to the surface-normal direction is a possible explanation for this absence.  That is, relaxation of the film may result in the presence of defects that inhibit the formation of charge and orbital ordering.  We note that a similar effect has been observed in Pr$_{0.5}$Ca$_{0.5}$MnO$_3$ films, as characterized by the reduced magnitude of the melting field.\cite{yang}

In the two other films--- those grown on LAO and NGO--- peaks with wavevectors consistent with the presence of charge and orbital ordering as exhibited by the bulk material are observed.  In both of these films, the correlation length of the peak associated with orbital order, and the transition temperature that is determined by this peak's disappearance, are reduced compared to the bulk material.  With respect to the peaks at the charge ordering wavevector, the two films behave differently than the bulk material, in that scattering is observed to persist above the transition temperature.  In fact in the film grown on LAO, a decoupling of this high-temperature scattering and that due to the charge order is clearly evident from measurements of the energy-dependence of the peak intensity, which exhibits a dramatic change at the transition temperature.  The two films also behave differently from each other at this wavevector, in that no resonant enhancement that would be indicative of charge ordering is observed in the film grown on NGO.

A question that follows from these results is:  do the films grown on LAO and NGO exhibit the charge and orbital ordering that is observed in the bulk material?  The answer to this question appears to be yes for the film grown on LAO.  While the different crystallographic symmetry of the film compared to the bulk material appears to lead to additional scattering at the charge order wavevector, which results in the different high-temperature behavior of the peak at the charge order wavevector, the low-temperature ordering wavevectors, resonant enhancements, and temperature dependences of the charge and orbital order peaks are strikingly similar to those observed in the bulk material.  In this film, the reduced orbital order correlation length and transition temperature can be attributed to the presence of a temperature-dependent, substrate-induced strain, which is strong enough to affect the behavior of the Bragg peaks.  Unfortunately, additional studies of the resonant scattering--- specifically, polarization- and azimuthal-dependence studies, which might further identify this ordering with that observed in the bulk--- are not feasible due to the thickness and orientation of the film.  Nevertheless, the similarity between the bulk material and this film remains compelling.  

For the film grown on NGO, the above question is more difficult, but we believe that the answer appears to be no.  This is based primarily on the fact that no resonant enhancement at either the charge or orbital order wavevectors was observed.  Near the (220) Bragg peak, the lack of a resonant enhancement at the orbital order wavevector can be attributed to the fact that scattering from a primarily transverse lattice distortion will dominate any resonant contribution, but the absence at the charge order peaks along (0{\it K}0) cannot be so easily explained.  In both the bulk material and the film grown on LAO, the resonant enhancement at the charge order wavevector results in a scattering intensity more than three times than that of the nonresonant scattering, as measured for incident photon energies below the Mn K edge.\cite{zim_long}  The lack of an observed resonant enhancement is therefore not due to a lack of intensity (the nonresonant scattering was $>$1000 s$^{-1}$), but appears to be an intrinsic property of the film.  

Another piece of evidence against bulk-like ordering being established in the film grown on NGO is the temperature-dependent behavior of the correlation length associated with the peak at the orbital order wavevector.  This can be seen in Figure 5, which contrasts the fitted values of the intensities and widths of the (2 1.5 0) peaks for the films grown on LAO and NGO.  Focusing on the widths, the behavior of the peak for the film grown on LAO is indicative of a transition in that the width appears to diverge at the transition temperature.  In contrast, for the film grown on NGO, the width of the peak is only weakly temperature-dependent and does not appear to broaden as the peak intensity diminishes and the peak ultimately disappears.  This behavior is, in fact, more reminiscent of the short-range correlations that are observed in manganite systems that exhibit no ordering in their low-temperature ground states--- for example, in the paramagnetic insulating phase of bulk La$_{0.7}$Ca$_{0.3}$MnO$_3$.\cite{dai,adams,me}  This observation suggests that the tendency to exhibit charge and orbital order exists in the film, but competing interactions--- perhaps related to the tensile, substrate-induced strain--- inhibit the transition into an ordered phase.  The possible presence of incommensurate peaks, which was discussed above, would add additional information to this picture.  That is, incommensurate peaks could suggest the presence of ordered discommensurations or an additional phase that is characterized by an incommensurate modulation. 

Finally, we note the interesting drop in intensity of the peaks at the bulk orbital order wavevector at low temperatures in both films (see Figure 5(a)).  While the intensity of a peak at this wavevector in the bulk material is observed to increase gradually with decreasing temperature and then remain constant below $\sim$100 K,\cite{zim_long} the intensity of the peaks in the two films reaches a maximum value around a temperature of 60 K before it {\it decreases} with decreasing temperature.  The explanation for this behavior is unclear, but a possible connection between it and the magnetization will be discussed in the next section.

\section{Resistivity and Magnetization Measurements of Films}

The resistivities of the Pr$_{0.6}$Ca$_{0.4}$MnO$_3$ films grown on LAO, NGO, and STO were measured at the University of Maryland using the conventional four-probe method, and the resulting data are displayed in Figure 6.  All films are observed to be insulating over the temperature range of 100--300 K, as is also the case in the bulk material.\cite{tomioka}  In terms of the magnitude of the resistivities, the film grown on NGO, which experiences the least amount of nominal and measured substrate-induced strain, is found to behave most like the bulk material.  

The onset of charge and orbital ordering can be identified in the bulk material by a change in slope in the measured resistivity.  In two of the films, subtle changes can be observed at temperatures of $\sim$230 and $\sim$270 K--- for the films grown on LAO and NGO, respectively.  We note that both of these temperatures exceed the temperatures at which the peaks at the bulk orbital order wavevector disappear in the respective films.

The DC magnetization of the Pr$_{0.6}$Ca$_{0.4}$MnO$_3$ films grown on LAO and STO were measured using a SQUID magnetometer, also at the University of Maryland (no magnetization data for the film grown on NGO could be obtained due to the large moment of the NGO substrate).  For each film, two sets of data were collected--- under zero-field-cooling (ZFC) and under field-cooling (FC), where the latter was in a field of 100 G applied in the plane of the film.  These data are shown in Figure 7.  In both films, the data indicate a reduced N$\acute{e}$el temperature compared to that observed in the bulk material (for which T$_N\approx$140 K).  In addition, the presence of a ferromagnetic component is suggested by the finite moment measured at low temperatures in both films.  This component could arise either from spin canting, which is believed to occur in the bulk material,\cite{tomioka} or from phase separation involving both ferromagnetic and antiferromagnetic phases.

Focusing now on the data from the film grown on LAO, we note the similar temperature dependences of the ZFC magnetization and the scattering intensity at the orbital order wavevector (see Figure 5(a)).  Intriguingly, it appears that the onset of antiferromagnetic ordering at a temperature of $\sim$60 K is correlated with a change in the orbital ordering in this film.  Since the correlation length of the peak at the orbital order wavevector exhibits  only a small change throughout this temperature regime (there is less than a 10\% difference in size of the ordered regions between the temperatures of 10 and 120 K), the decrease in scattering intensity indicates a decrease in the magnitude of the lattice distortion at the orbital order wavevector and/or a reduction in the number of ordered regions.  Therefore, antiferromagnetic ordering appears to coincide with a destabilization of the orbital ordering in this film--- a result that is the converse to what has been observed in bulk LaMnO$_3$\cite{murakami2} and Pr$_{0.5}$Ca$_{0.5}$MnO$_3$.\cite{zim_long}  The explanation for this interesting effect is unclear, and should be the focus of further investigations.

Turning now to the FC data for the film grown on LAO, a slight hysteresis can be observed (see the inset to Figure 7).  In bulk manganites, such hysteresis has been attributed to the presence of disorder\cite{moreo}--- for example, in the distribution of the Pr and Ca cations--- and this could also apply to the film.  An additional source of disorder that is unique to the film could be the effects of substrate-induced strain.  Specifically, relaxation of the film could exceed the elastic threshhold and result in the presence of defects.

Finally, in the film grown on STO, the ferromagnetic component is observed to be much larger than in the film grown on LAO.  In the phase separation picture mentioned above, this result could explain the absence of charge and orbital ordering in this film since a ferromagnetic phase would compete with a phase exhibiting charge and orbital order.  A larger ferromagnetic component would therefore arise at the expense of charge and orbital ordering.  We note that this explanation is also consistent with the different expected growth modes for the two films.  That is, the island growth mode that is expected to occur under compressive strain has been found through magnetization measurements to disfavor the ferromagnetic metallic phase in La$_{0.67}$Ca$_{0.33}$MnO$_3$ films.\cite{amlan}

\section{Summary}

Substrates providing compressive (LAO) and different amounts of tensile (NGO and STO) strain were used to study the effects of substrate-induced strain on Pr$_{0.6}$Ca$_{0.4}$MnO$_3$ films.  In all three films, the room temperature, crystallographic symmetry was observed to be different from that of the bulk material.  Specifically, the films grown on LAO and NGO were found to be monoclinic, with the substrate-induced strain primarily affecting the values of the in-plane lattice constant, {\it c}, and the angle between the out-of-plane, {\it a} and {\it b}, axes.  In these two films, the in-plane lattice constant was observed to be nearly clamped to the substrate, resulting in a negligible lattice mismatch between the film and substrate at room temperature.  In contrast, in the film grown on STO, the in-plane lattice constant differed only slightly relative to the bulk value, and the film was observed to be tetragonal.  The substrate-induced strain in this film also appeared to undergo relaxation within the film, as evidenced by the broad peaks observed in longitudinal scans.  Perhaps not surprisingly, given that this relaxation may result in the presence of defects, no low-temperature ordering was observed in this film.  We also note in this context that recent work suggests that correlations of the type seen in the film on NGO are suppressed in samples with tetragonal symmetry.\cite{valery2}  Thus it is possible that the symmetry of the film on STO also plays a role in the suppression of any correlations.

In the film grown on LAO, two sets of peaks with wavevectors, resonant enhancements, and temperature dependences similar to those associated with charge and orbital ordering in the bulk material were observed at low temperatures, although both the transition temperature and orbital order correlation length were found to be reduced.  One explanation for this reduction could lie in the increasing strain with decreasing temperature that results from the different temperature dependences of the film and substrate lattice constants, which was observed to affect the correlation length of the film Bragg peaks--- broadening them with decreasing temperature.  In this film, we also observed a connection between the magnetization and the scattering intensity due to the lattice distortion associated with the orbital ordering.  The onset of antiferromagnetic ordering at a temperature of $\sim$60 K appears to coincide with a destabilization of the orbital ordering, apparently resulting in a decrease in the magnitude of the lattice distortion and/or a reduction in the number of ordered regions.

In the film grown on NGO, broad peaks at the bulk orbital order wavevector were observed at low temperatures.  However, the lack of resonant enhancements at the charge and orbital order wavevectors, as well as the weak temperature dependence of the correlation length of the peaks at the orbital wavevector, suggest that the film does not exhibit bulk-like charge and orbital order.  More likely, a tendency to order produces short-range correlations that result in the broad peaks at the bulk orbital order wavevector, but a transition into an ordered phase never occurs.

Considering that of the three films investigated, the film grown on NGO experiences the least amount of nominal and measured, substrate-induced strain, the lack of bulk-like ordering in this film is a surprise.  The charge and orbital ordering observed in bulk Pr$_{0.6}$Ca$_{0.4}$MnO$_3$ appears to be more robust under compressive, rather than tensile, strain.  If one considers the effects of substrate-induced strain on the structure of the material, a possible explanation presents itself:  the different effects of compressive and tensile strain on the Mn--O--Mn bond angles.  That is, compressive (tensile) strain results in two out of the three bond angles being decreased (increased), which decreases (increases) the bandwidth.  A larger bandwidth destabilizes the charge and orbital ordering, which is therefore more robust in this material in the presence of a larger nominal, compressive strain than a smaller nominal, tensile strain.  Note that the complementary effect--- the stabilization of a ferromagnetic metallic state under tensile strain--- has been reported in La$_{1-x}$Ba$_x$MnO$_3$ films.\cite{yuan}

In conclusion, substrate-induced strain caused by a lattice mismatch between Pr$_{0.6}$Ca$_{0.4}$MnO$_3$ and LAO, NGO, and STO substrates is observed to have dramatic effects on several properties--- including the crystallographic symmetry, charge and orbital ordering, and magnetic behavior--- of Pr$_{0.6}$Ca$_{0.4}$MnO$_3$ films.  These results underline the importance of carrying out thorough studies of the structure, low-temperature charge and orbital ordering, transport, and magnetization in any effort with an ultimate goal involving manganite film applications.

The work at Brookhaven, both in the Physics Department and at the NSLS, was supported by the U.S. Department of Energy, Division of Materials Science, under Contract No. DE-AC02-98CH10886.  The work at the University of Maryland was supported by the National Science Foundation MRSEC (DMR-00-8008).  The work at the CMC beamlines was supported, in part, by the Office of Basic Energy Sciences of the U.S. Department of Energy and by the National Science Foundation, Division of Materials Research.  Use of the Advanced Photon Source was supported by the Office of Basic Energy Sciences of the U.S. Department of Energy under Contract No. W-31-109-Eng-38.  We thank A. Sharma for her contribution in the preparation of the films.

\clearpage

\begin{table}
\caption{substrate-induced strain}
\begin{ruledtabular}
\begin{tabular}{ccc}
& nominal ($\frac{c_{substrate}-c_{bulk}}{c_{bulk}}$) & measured ($\frac{c_{film}-c_{bulk}}{c_{bulk}}$) \\  \hline
Pr$_{0.6}$Ca$_{0.4}$MnO$_3$ on LAO & -1.4 \% & -1.4 $\pm$ 0.3 \% \\
Pr$_{0.6}$Ca$_{0.4}$MnO$_3$ on NGO & 0.6 \% & 0.3 $\pm$ 0.3 \% \\
Pr$_{0.6}$Ca$_{0.4}$MnO$_3$ on STO & 1.9 \% & 0.5 $\pm$ 0.3 \% \\
\end{tabular}
\end{ruledtabular}
\end{table}

\begin{table}
\caption{room temperature lattice parameters}
\begin{ruledtabular}
\begin{tabular}{ccccc}
& a ($\AA$) & b ($\AA$) & c ($\AA$) & $\gamma$ ($^{\circ}$) \\  \hline
bulk Pr$_{0.6}$Ca$_{0.4}$MnO$_3$\footnotemark[1] & 5.415 & 5.438 & 7.664 & 90 \\
Pr$_{0.6}$Ca$_{0.4}$MnO$_3$ on LAO & 5.43 $\pm$ 0.01 & 5.44 $\pm$ 0.01 & 7.56 $\pm$ 0.02 & 88.2 $\pm$ 0.1 \\
Pr$_{0.6}$Ca$_{0.4}$MnO$_3$ on NGO & 5.42 $\pm$ 0.01 & 5.42 $\pm$ 0.01 & 7.69 $\pm$ 0.02 & 91.4 $\pm$ 0.1 \\
Pr$_{0.6}$Ca$_{0.4}$MnO$_3$ on STO & 5.44 $\pm$ 0.01 & 5.44 $\pm$ 0.01 & 7.70 $\pm$ 0.02 & 90.0 $\pm$ 0.1 \\
\end{tabular}
\end{ruledtabular}
\footnotetext[1]{values for bulk material taken from Jirak {\it  et al.} \cite{jirak}}
\end{table}
 
\begin{figure}
\includegraphics[width=7cm]{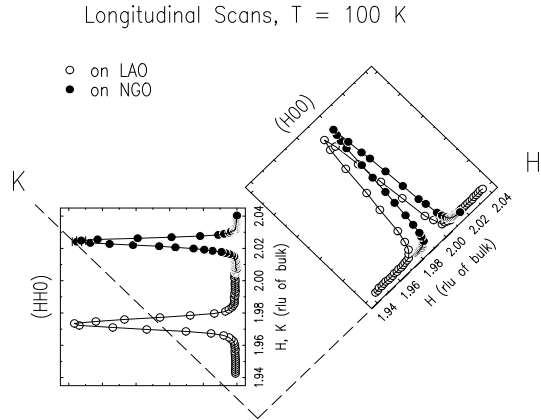}
\caption{Longitudinal scans through the (220) and (200) peaks for films grown on LAO ($\circ$) and NGO ($\bullet$).}
\end{figure}

\begin{figure}
\includegraphics[width=5cm]{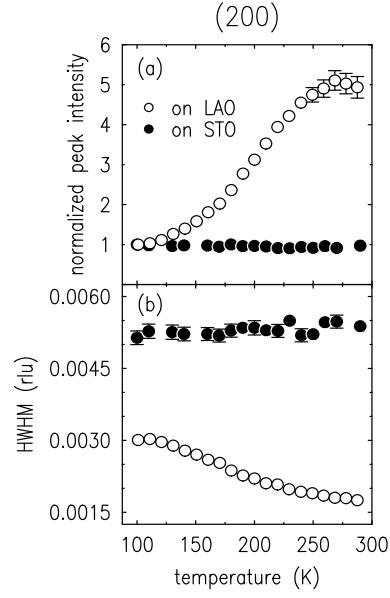}
\caption{Fitted values for the (200) Bragg peak intensities (a) and widths (b), for films grown on LAO ($\circ$) and STO ($\bullet$).  The peak intensities are normalized to equal 1 at a temperature of 100 K.}
\end{figure}

\begin{figure}
\includegraphics[width=6cm]{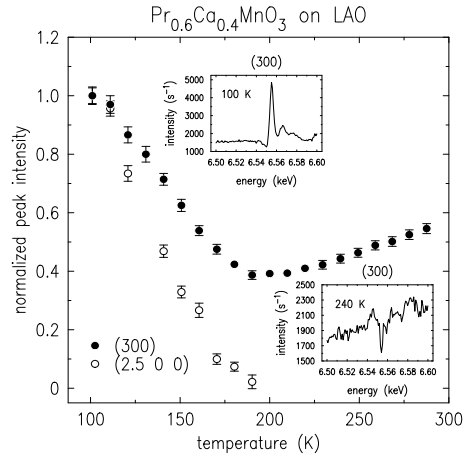}
\caption{Fitted values for the (300) charge ($\bullet$) and (2.5 0 0) orbital ($\circ$) order peak intensities, normalized to equal 1 at a temperature of 100 K.  Insets show energy scans carried out at the (300) peak at temperatures of 100 and 240 K.}
\end{figure}

\begin{figure}
\includegraphics[width=5cm]{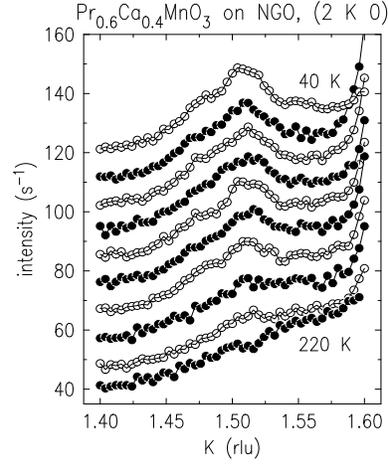}
\caption{Reciprocal space scans through the (2{\it K}0) peak, measured as a function of temperature, in steps of 20 K.  For clarity, each data set is shifted upward by 10 s$^{-1}$ with respect to the next higher temperature data set.}
\end{figure}

\begin{figure}
\includegraphics[width=5cm]{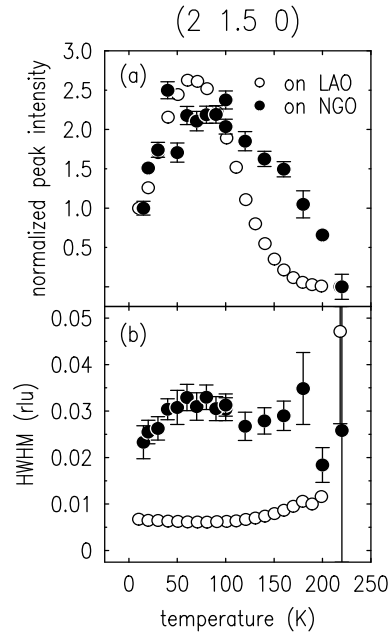}
\caption{Fitted values for the (2 1.5 0) peak intensities (a) and widths (b), for films grown on LAO ($\circ$) and NGO ($\bullet$).  The peak intensities are normalized to equal 1 at the lowest temperature studied.}
\end{figure}

\begin{figure}
\includegraphics[width=6cm]{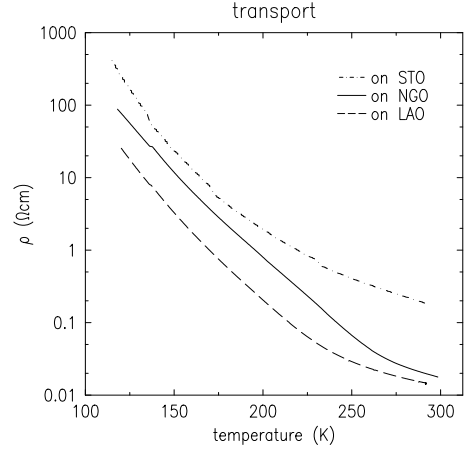}
\caption{Resistivity as a function of temperature for films grown on STO (-- $\cdot$ --), NGO (---), and LAO (-- --).}
\end{figure}

\begin{figure}
\includegraphics[width=6cm]{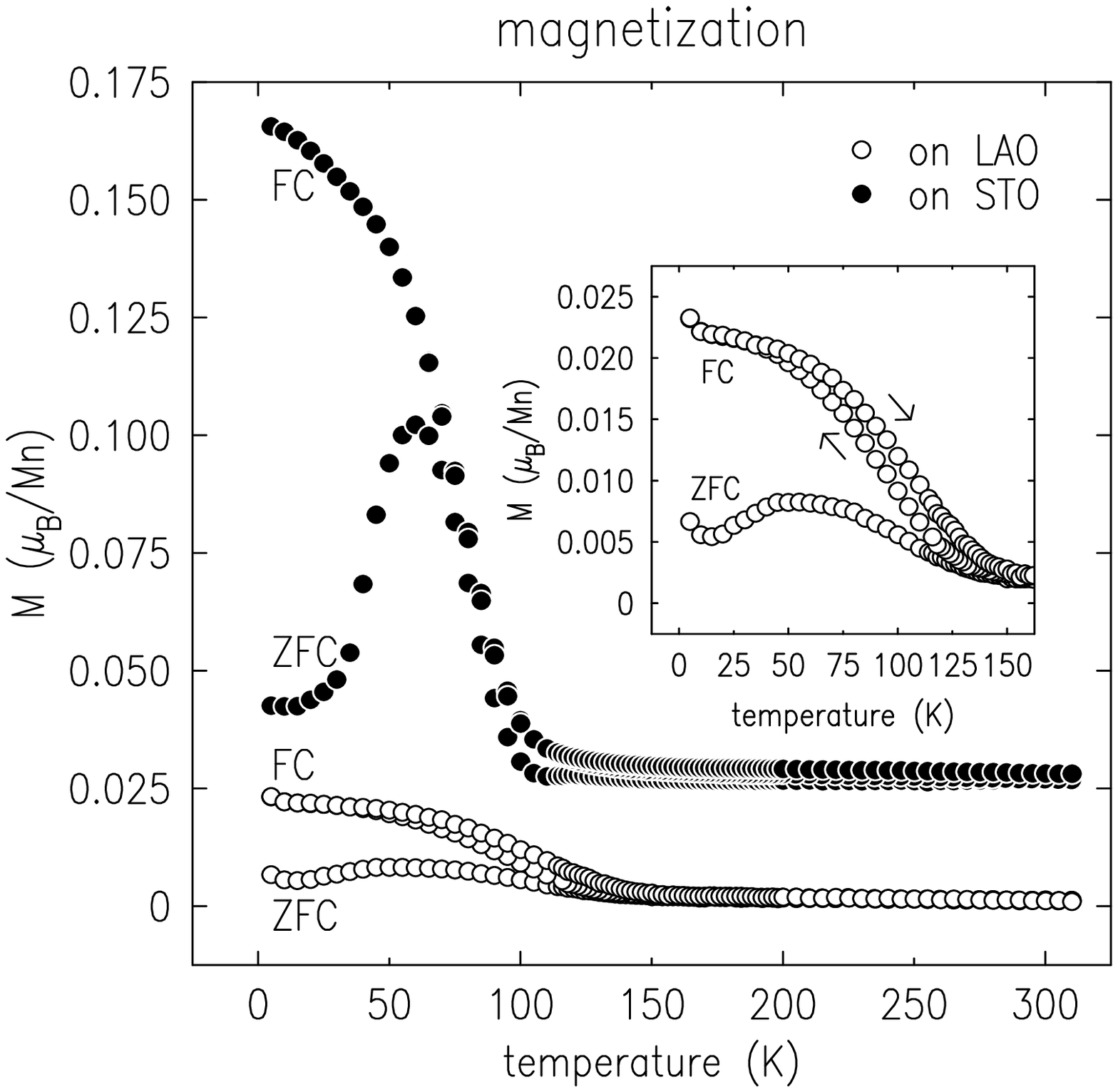}
\caption{Magnetization as a function of temperature for films grown on LAO ($\circ$) and STO ($\bullet$).  ZFC and FC refer to zero-field-cooling and field-cooling (in a field of 100 G applied in the plane of the film), respectively.  Inset shows a blow-up of the low-temperature data for the film grown on LAO, and the arrows indicate the direction in which the temperature was swept while the data were collected.}
\end{figure}


\begin{thebibliography}{99}
\bibitem{rutherford} Note that the stoichiometry of the films was confirmed using Rutherford backscattering spectrometry measurements on other films that were grown from the same pellet.
\bibitem{3headed} Note that the mosaic widths of the three (220) peaks of the film grown on STO varied in the range of 0.1--0.3$^{\circ}$.  The possible origin of the three peaks is described in section III.
\bibitem{caveat} Note that for all of the measurements described in this section, care was initially taken to avoid x-ray induced effects during the search for low-temperature charge and orbital ordering.  But as in the film grown under compressive strain on LAO, no evidence of x-ray induced effects were observed.
\end{thebibliography}
\end{document}